\documentclass[twocolumn ,showpacs
 ,secnumarabic, nobibnotes,  nofootinbib, aps, prd]{revtex4}
\usepackage{amsmath}
\usepackage{amssymb}
\usepackage{bm}
\usepackage{graphicx}

\begin{document}

\title{Generalized Cahn effect and parton 3D motion in {a} covariant
approach }
\author{Petr Zavada}
\email{zavada@fzu.cz}
\affiliation{Institute of Physics AS CR, Na Slovance 2, CZ-182 21 Prague 8, Czech Republic}

\begin{abstract}
The Cahn effect and the unintegrated unpolarized parton distribution
function $f_{1}^{q}(x,\mathbf{p}_{T})$ are studied in a covariant approach. {%
The} Cahn effect is compared with some other effects due to the parton
intrinsic motion. The comparison suggests that {the} present understanding
of parton transverse momenta and intrinsic motion in general is still rather
incomplete. The new {relation} for $f_{1}^{q}(x,\mathbf{p}_{T})$ is obtained
in the framework of the covariant parton model {from} which a prediction for
this distribution function follows.
\end{abstract}

\maketitle

\section{Introduction}

\label{sec1}Studies of the transverse momentum dependent (or 'unintegrated')
parton distribution functions (TMDs) \cite{tmds} open {a} new way to {a}
better understanding of the partonic quark-gluon structure of the nucleon.
At the same time it is evident, that {the} explanation of some experimental
observations could be hardly possible without {a} more accurate and
realistic 3D picture of the nucleon, which naturally includes transverse
motion. The azimuthal asymmetry in {the} distribution of hadrons produced in
deep-inelastic lepton-nucleon scattering (DIS), known as the Cahn effect 
\cite{Cahn:1978se, Cahn:1989yf}, is a classical example. The role of the
quark (transversal) intrinsic motion is crucial {also} for {the} explanation
of some spin effects, like the asymmetries in particle production related to
the direction of proton polarization \cite{Airapetian:1999tv,
Airapetian:2001eg, Airapetian:2004tw, Adams:1991rw, Adams:1991cs,
Bravar:1996ki, Adams:2003fx}.

In our previous study we proposed {a} covariant parton model, in which the
3D picture of parton momenta with rotational symmetry in the nucleon rest
frame represents a basic input \cite{Zavada:1996kp, Zavada:2001bq,
Zavada:2002uz, Efremov:2004tz, Zavada:2007ww, Efremov:2008mp, Efremov:2009ze}%
. At the same time the model is based on the assumption, that for
sufficiently {large} momentum transfer $Q^{2}$, the quarks can be considered
as almost free due to the asymptotic freedom. It appears, that the main
potential of this approach is {the} implication of some old and new sum
rules and relations among various parton distribution functions. The sum
rules which relate {the} structure functions $g_{1}$ and $g_{2}$ {in a
Wandzura}-Wilczek {approximation} and some others are proved in \cite%
{Zavada:2001bq}. Assuming the $SU(6)$ symmetry (in addition to the
covariance and rotational symmetry) we have proved relations between
polarized and unpolarized structure functions \cite{Zavada:2002uz}, which
agree very well with the experimental data. In \cite{Efremov:2004tz} we
studied transversity in the framework of this model and we {derived
relations between} transversity and helicity. Recently, we generalized the
model to include also the pretzelosity distribution \cite{Efremov:2008mp}
and {derived relations which connect} helicity, transversity and
pretzelosity. Finally, with the same model we studied the TMDs and a set of
relations among them \cite{Efremov:2009ze}. Further, in {the} framework of
the model we demonstrated that the 3D picture of parton momenta inside the
nucleon is a necessary input for {a} consistent accounting for quark orbital
angular momentum (OAM). {The dominanting contribution} of the OAM {to} the
nucleon spin is a consequence of the quark relativistic motion inside the
nucleon, i.e.\ when quark $mass\ll momentum$ in the nucleon rest frame. In
this case only the total angular momentum $J_{z}^{q}=L_{z}^{q}+S_{z}^{q}$\
is {a} good quantum number and we obtained mean values of the quark orbital
and spin components: $\left\langle L_{z}^{q}\right\rangle =$ $2\left\langle
S_{z}^{q}\right\rangle =\Delta \Sigma $ or $\left\langle
J_{z}^{q}\right\rangle =\left\langle S_{z}^{q}\right\rangle +\left\langle
L_{z}^{q}\right\rangle =\frac{{3}}{2}\Delta \Sigma $ \cite{Zavada:2007ww}.

{A} comparison of {the} obtained relations and predictions with experimental
data is very important and interesting from phenomenological point of view.
It allows to judge to which extent the experimental observation can be
interpreted in terms of simplified, intuitive notions. The obtained picture
of the nucleon can be {a} useful complement to the exact but more
complicated theory of the nucleon structure based on the QCD. For example,
the covariant parton model can be {a} useful tool for separating {QCD effects%
} from effects of relativistic kinematics.

In this work we study further aspects of {the} intrinsic motion of quarks. {%
The} Cahn effect is due to transverse momentum of quarks and in Sec. \ref%
{sec2}. we analyze the conditions, which induce this effect in more detail.
We show, that the azimuthal asymmetry is a general consequence of {the}
intrinsic motion of constituents inside the composite target. We obtain the
corresponding asymmetry {as function of parton} transverse momentum in {the}
covariant approach. In Sec. \ref{sec3}. we make a comparison of the data on
average transverse {momenta} of the quarks obtained by the method based on
the Cahn effect with the data obtained by {other also} model-dependent
methods. In Sec. \ref{sec4}. we analyze the unpolarized TMD defined in our
previous study \cite{Efremov:2009ze} and as a result we obtain the relation
between this unintegrated distribution and its integrated counterpart. This {%
relation} allows to make a prediction for the TMD using the known parton
distribution function. We also make a detailed comparison with the recent
approach by U.D'Alesio, E.Leader and F.Murgia \cite{D'Alesio:2009kv}, in
which an equivalent {relation} has been obtained. Finally in Sec. \ref{sec5}%
. we summarize {the} obtained results.

\section{Cahn effect: manifestation of the intrinsic motion}

\label{sec2}The Cahn effect, which is related to azimuthal asymmetry of
struck quarks in DIS, is due to the nonzero transverse momentum of quarks
inside the nucleon. {The} probability $W=\left\vert M_{fi}\right\vert ^{2}$
of the elementary lepton-quark scattering in one photon exchange
approximation is given by the expression%
\begin{equation}
W(\hat{s},\hat{u})\varpropto \hat{s}^{2}+\hat{u}^{2},  \label{c1}
\end{equation}%
where {the} Mandelstam variables depend on the azimuthal angle $\varphi $
(angle between leptonic and hadronic planes) as:%
\begin{eqnarray}
\hat{s}^{2} &=&\frac{Q^{4}}{y^{2}}\left[ 1-4\frac{p_{T}}{Q}\sqrt{1-y}\cos
\varphi \right] +\mathcal{O}\left( \frac{p_{T}^{2}}{Q^{2}}\right) ,
\label{c2} \\
\hat{u}^{2} &=&\frac{Q^{4}}{y^{2}}\left( 1-y\right) ^{2}\left[ 1-4\frac{p_{T}%
}{Q}\frac{\cos \varphi }{\sqrt{1-y}}\right] +\mathcal{O}\left( \frac{%
p_{T}^{2}}{Q^{2}}\right) ,  \label{c3}
\end{eqnarray}%
where $p_{T}$\ is the quark momentum component transverse to the photon
momentum $\mathbf{q}$, $Q^{2}\equiv -q^{2}$ \cite{Anselmino:2005nn}.
Apparently, {the} dependence on $\varphi $ disappears for $p_{T}\rightarrow 0
$. {The} intrinsic motion of the constituents creating the composite target
is a necessary condition for {the} appearance of the effect. The Cahn effect
is a kinematical effect accompanying the QED scattering of leptons on quarks
inside the nucleon and its origin is different from that of the QCD
higher-twist effects \cite{Mulders:1995dh,Bacchetta:2006tn,Berger:1979xz}.
At the same time it is evident, that {the} intrinsic quark motion in itself
is due to {non-}pertubative QCD. {The} Mandelstam variables in terms of {the}
lepton and quark momenta ($l,p$) read%
\begin{eqnarray}
\hat{s} &=&\left( l+p\right) ^{2}=2pl+m_{l}^{2}+m_{q}^{2},  \label{c4} \\
\hat{u} &=&\left( p-l^{\prime }\right) ^{2}=-2pl+Q^{2}+m_{l}^{2}+m_{q}^{2},
\label{c5}
\end{eqnarray}%
where $m_{l},m_{q}$ are {the} corresponding masses. Obviously, one can
substitute the variables of the probability (\ref{c1}):%
\begin{equation}
\hat{s},\hat{u}\rightarrow pl,Q^{2};\qquad W(\hat{s},\hat{u})\rightarrow
W(pl,Q^{2}).  \label{c5a}
\end{equation}%
The probability $W$ expressed in terms of the new variables clearly
demonstrates {an} azimuthal symmetry of $\mathbf{p}$ with respect to the
lepton beam direction $\mathbf{l}$, which represents the axis of {the}
azimuthal symmetry. It follows, that the photon direction $\mathbf{q}$ being
different from the direction $\mathbf{l}$, in general cannot be the second
axis of {the} azimuthal symmetry. In fact, this is {the} essence of the Cahn
effect, see Fig. \ref{fg1}a. Let us consider two reference frames: 
\begin{figure}[tbp]
\includegraphics[width=9cm]{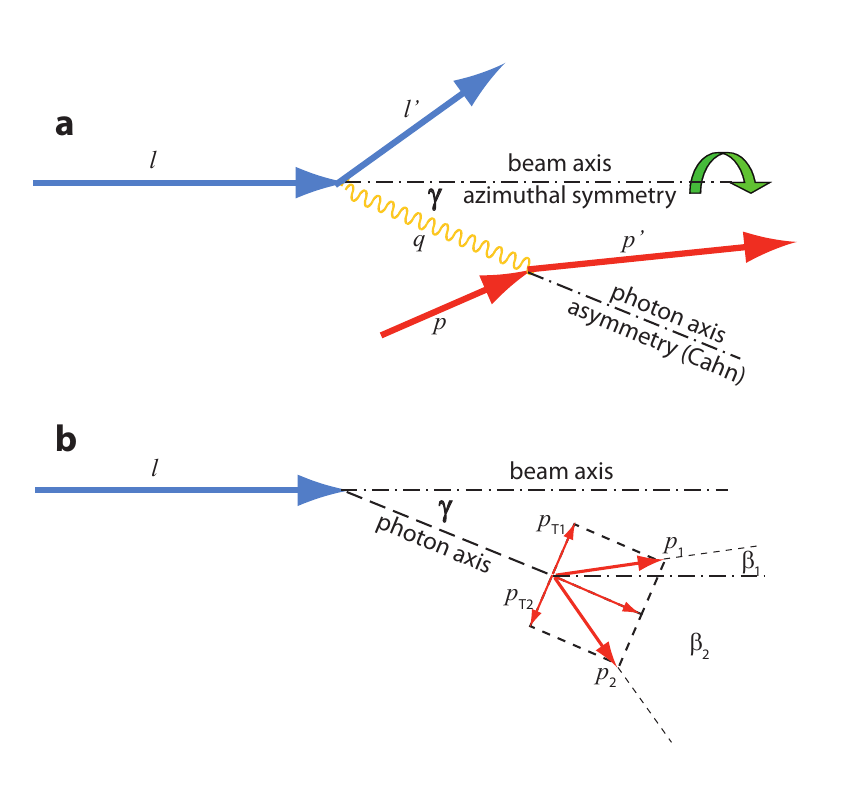}
\caption{{The} interaction of {a} lepton with {a} quark defines two axes of
different symmetry (\textbf{a}). {The} azimuthal asymmetry as a result of
variable collision energy (\textbf{b}), see text.}
\label{fg1}
\end{figure}

\textit{A}. The nucleon rest frame, where the first axis is directed along $%
\mathbf{q}$ and {the} projection of $\mathbf{l}$ on the plane transversal to 
$\mathbf{q}$\ defines {a} second axis. {The} azimuthal angle $\varphi $ and {%
the} transverse momentum $p_{T}$ are defined equally as above ($p_{T}$ and $%
\varphi $ do not change under any Lorentz boost along $\mathbf{q}$), so the
quark momentum $\mathbf{p}$ in this frame has the components: 
\begin{equation}
\mathbf{p}_{A}=(p_{1},p_{T}\cos \varphi ,p_{T}\sin \varphi ).  \label{c6}
\end{equation}

\textit{B}. The nucleon rest frame, where the first axis is directed along $%
\mathbf{l}$ and projection of $-\mathbf{q}$ on the plane transversal to $%
\mathbf{l}$ defines second axis. This reference frame is obtained by {a}
rotation of the frame \textit{A} by {an} angle $\gamma $ around {a} third
axis, so the quark momentum has the new components:%
\begin{eqnarray}
\mathbf{p}_{B} &=&(p_{1}\cos \gamma -p_{T}\sin \gamma \cos \varphi ,
\label{c7} \\
&&p_{T}\cos \gamma \cos \varphi +p_{1}\sin \gamma ,p_{T}\sin \varphi ). 
\notag
\end{eqnarray}%
The angle $\gamma $\ is defined as%
\begin{equation}
\cos \gamma =\frac{q_{L}}{\left\vert \mathbf{q}\right\vert },\qquad \sin
\gamma =\frac{q_{T}}{\left\vert \mathbf{q}\right\vert },  \label{c8}
\end{equation}%
where $q_{L}$ and $q_{T}$ are {the} longitudinal and transversal components
of the photon momentum in the frame \textit{B}, $\mathbf{q}_{B}=\left(
q_{L},q_{T},0\right) $. For {the} lepton energy $l_{0}$ ({the} lepton mass
will be neglected in the {following}) one can obtain \cite{Zavada:1996kp}:%
\begin{equation}
\frac{\left\vert q_{L}\right\vert }{\nu }=1+\frac{M}{l_{0}}x_{B},\qquad 
\frac{\left\vert \mathbf{q}\right\vert }{\nu }=\sqrt{1+\frac{4M^{2}}{Q^{2}}%
x_{B}^{2}}  \label{c9}
\end{equation}%
and

\begin{equation}
\frac{q_{T}}{\nu }=\sqrt{\left( \frac{4M^{2}}{Q^{2}}-\frac{M^{2}}{l_{0}^{2}}%
\right) x_{B}^{2}-\frac{2M}{l_{0}}x_{B}},  \label{c10}
\end{equation}%
where the standard notation is used:%
\begin{equation}
x_{B}=\frac{Q^{2}}{2M\nu },\qquad \nu =l_{0}-l_{0}^{\prime }.  \label{c11}
\end{equation}%
Now the variable $pl$ can be expressed as%
\begin{equation}
pl=\left( p_{0}-p_{1}\cos \gamma -p_{T}\sin \gamma \cos \varphi \right)
l_{0}.  \label{c12}
\end{equation}%
This variable, after inserting {into the} relations (\ref{c4}), (\ref{c5})
allows to exactly calculate {the} azimuthal dependence of the probability (%
\ref{c1}).

If one assumes%
\begin{equation}
Q^{2}\gg 4M^{2}x_{B}^{2},\qquad l_{0}\gg Mx_{B},  \label{c12a}
\end{equation}%
then the relations (\ref{c8}) and (\ref{c9}) give%
\begin{equation}
\left\vert \mathbf{q}\right\vert \approx \left\vert q_{L}\right\vert \approx
\nu ,\qquad \cos \gamma \approx 1.  \label{c12b}
\end{equation}%
Now, since%
\begin{equation}
p_{1}=\frac{\mathbf{pq}}{\left\vert \mathbf{q}\right\vert }=\frac{p_{0}\nu
-pq}{\left\vert \mathbf{q}\right\vert },  \label{c12c}
\end{equation}%
the relation (\ref{c12}) is modified as%
\begin{equation}
pl\approx \left( \frac{pq}{\nu }-\frac{p_{T}q_{T}}{\nu }\cos \varphi \right)
l_{0}.  \label{c12d}
\end{equation}%
Further, Eq. (\ref{c10}) is rearranged as%
\begin{equation}
\frac{q_{T}}{\nu }=\frac{2Mx_{B}}{Q}\sqrt{1-\frac{\nu }{l_{0}}-\frac{Q^{2}}{%
4l_{0}^{2}}}.  \label{c12e}
\end{equation}%
Since the complete expression for the probability $W(\hat{s},\hat{u})$
involves the $\delta${--}function term%
\begin{equation}
\delta \left( \left( p+q\right) ^{2}-m_{q}^{2}\right) =\delta \left(
2pq+q^{2}\right) =\frac{1}{2Pq}\delta \left( \frac{pq}{Pq}-x_{B}\right) ,
\label{c12f}
\end{equation}%
where $P$ is {the} nucleon momentum, one can replace {the} product $pq$ in (%
\ref{c12d}) by $Mx_{B}\nu $. Then, assuming $4l_{0}^{2}\gg Q^{2}$, after
inserting (\ref{c12e}) into (\ref{c12d}) one gets%
\begin{equation}
pl\approx \frac{Q^{2}}{2y}\left( 1-\frac{2p_{T}\sqrt{1-y}}{Q}\cos \varphi
\right) ,\qquad  \label{c12g}
\end{equation}%
where%
\begin{equation*}
y=\frac{\nu }{l_{0}}=\frac{Pq}{Pl},\qquad \frac{Q^{2}}{2y}=x_{B}Pl. 
\end{equation*}%
Now, the term%
\begin{equation}
\lambda =\frac{2p_{T}\sqrt{1-y}}{Q}\cos \varphi  \label{c12h}
\end{equation}%
represents a "small" correction and one can check, that {the} Mandelstam
variables (\ref{c4}),(\ref{c5}) in which the term $pl$ is replaced by the
expression (\ref{c12g}) and {the} quark masses are neglected, give the
relations (\ref{c2}),(\ref{c3}).

Now the probability $W(pl,Q^{2})$ can be expanded as 
\begin{gather}
W(pl,Q^{2})=\left. W(pl,Q^{2})\right\vert _{\lambda =0}-\left. \frac{%
\partial W(pl,Q^{2})}{\partial \left( pl\right) }pl\right\vert _{\lambda
=0}\lambda +...  \notag \\
\approx \left. W(pl,Q^{2})\right\vert _{\lambda =0}\left( 1-\left. \frac{%
\partial \ln W(pl,Q^{2})}{\partial \ln (pl)}\right\vert _{\lambda =0}\lambda
\right) .  \label{c12i}
\end{gather}%
Let us make some remarks on this relation:

\textit{i)} The relation implies, that {the} azimuthal asymmetry of the
recoiled quark is described by the distribution%
\begin{equation}
P(\varphi )=\left( 1-a\cos \varphi \right) ,  \label{d1}
\end{equation}%
where%
\begin{equation}
a=\frac{2\sqrt{1-y}}{Q}\cdot \left[ \frac{\partial \ln W(pl,Q^{2})}{\partial
\ln (pl)}\right] _{\lambda =0}\cdot \left\langle p_{T}\right\rangle .
\label{d2}
\end{equation}%
From the analysis of experimental data one can obtain the parameter $a$.
Obviously for obtaining $\left\langle p_{T}\right\rangle $ one has to know
also the term involving differentiation of $W$. This term can be estimated
either from the model (Eq. (\ref{c1})) or from the experiment, if the data
for a few lepton energies are available.

\textit{ii)} {The} azimuthal asymmetry generated by the probability $%
W(pl,Q^{2})$ has {a} simple geometrical interpretation. In Fig. \ref{fg1}b {%
the} two momenta $\mathbf{p}_{1},\mathbf{p}_{2}$ with opposite transverse
components $\mathbf{p}_{T1},\mathbf{p}_{T2}$ correspond to different
collision energies $\hat{s}_{1},\hat{s}_{2}$ since%
\begin{equation}
\hat{s}=2pl=2\left( p_{0}l_{0}-\left\vert \mathbf{p}\right\vert \left\vert 
\mathbf{l}\right\vert \cos \beta \right) ,  \label{c12k}
\end{equation}%
where $\beta $\ is angle between the lepton and quark momenta. Obviously $%
\hat{s}_{1}<\hat{s}_{2}$ in this figure and because $W$ depends on $\hat{s}$%
, then the two corresponding momenta $\mathbf{p}_{1},\mathbf{p}_{2}$ give
different probabilities. In this way the asymmetry is generated. The figure
reflects {the} necessary conditions for the asymmetry:%
\begin{equation}
\sin \gamma >0,\qquad \frac{\partial W}{\partial s}>0,\qquad \left\langle
p_{T}\right\rangle >0,  \label{d3}
\end{equation}%
which correspond to the three factors in {the} asymmetry parameter (\ref{d2}%
).

\textit{iii)} In fact we have shown, that this asymmetry can be expected in 
\textit{any} process $l+p\rightarrow l^{\prime }+p^{\prime }$ described by
the probability $W(\hat{s},\hat{u})$, which is defined only by the incoming
particle vector $l$, momentum transfer $q$ and by the parton vector $p$ (or
another constituent of composite target having some distribution of
intrinsic $\mathbf{p}_{T}$).

In our case the probability $W$ is related to the individual lepton-quark
scattering, which is only one stage of the Cahn effect. For complete
phenomenology of the effect in lepton-nucleon DIS one needs further inputs:

\textit{a)} 3D distribution $G(p)d^{3}p$ of parton momenta in the nucleon.
The covariant approach will be studied in Sec.~\ref{sec4}.

\textit{b)} Fragmentation of recoiled quark and transfer of azimuthal
asymmetry to hadrons. It is a complex stage {containing} both pertubative
and {non-}pertubative QCD aspects, {but some standard parameterization of
the fragmentation function can be used, like in \cite{Anselmino:2005nn}.}

\section{What do we know about intrinsic motion?}

\label{sec3}In the lepton-quark scattering the distribution $G(p)$ controls
the distributions of momenta of the scattered lepton and the recoiled quark.
And vice versa, from the knowledge of the distributions of the scattered
leptons or quarks (in real analysis hadrons from the quark fragmentation),
one can obtain information about the initial distribution by two independent
ways. {The} comparison of the results can serve as a consistency check. So
we can analyze two sets of data:\medskip

\textit{A. Leptonic data}

The nucleon structure function $F_{2}(x,Q^{2})$ is obtained {from} the
analysis of lepton data from DIS experiments.

\textit{i)} {The} interpretation of this function in {the} framework of the
usual, \textit{non-covariant} parton model suggests, that {(valence+sea)}
quarks carry approximately only 50\% of the nucleon energy-momentum. It
follows that the one valence quark can carry less than roughly 15\% \ (more
exactly $\left\langle x\right\rangle =\int xq_{val}(x)dx/\int
q_{val}(x)dx=0.155(0.118)$ for the \textit{u(d)} valence quarks at $%
Q^{2}=4GeV^{2}$ \cite{Martin:2004dh}). This {estimate} follows from the
approach in the nucleon infinite momentum frame, where the transversal
momentum of the quarks is neglected.

\textit{ii) }The analysis of the function $F_{2}(x,Q^{2})$ in the framework
of the \textit{covariant} parton models gives the following results. The
model \cite{Jackson:1989ph} gives the prediction for the dependence $%
\left\langle p_{T}^{2}/M^{2}\right\rangle $ on $x$: the ratio vanishes at $%
x=0$ and $\ x=1$ and reaches the peak value $0.04-0.05$ at $x\approx 0.5$. {A%
} very similar picture is obtained also in \cite{D'Alesio:2009kv}. {Since} $%
p_{T}/M\approx 0.2$ at the peak, the mean value {averaged} over $x$ must be
smaller. These results are quite consistent with those obtained in the
covariant model in which we obtained for massless quarks the relation \cite%
{Zavada:1996kp}%
\begin{equation}
p_{T}^{2}\leq M^{2}x\left( 1-x\right) \equiv {p_{T\max }^{2}}(x)  \label{c18}
\end{equation}%
and for average momentum of the valence quarks in the nucleon \textit{rest
frame} we get \cite{Zavada:2007ww}%
\begin{equation}
\left\langle \left\vert \mathbf{p}_{val}\right\vert \right\rangle \approx
0.1GeV,\qquad \left\langle p_{Tval}\right\rangle =\frac{\pi }{4}\left\langle
\left\vert \mathbf{p}_{val}\right\vert \right\rangle .  \label{c19}
\end{equation}

\textit{iii)} The statistical model \cite{Bhalerao:1999xy} of the nucleon
gives {a} very good description of the unpolarized ($F_{2}^{p,n}$) and
polarized ($g_{1}^{p,n}$) structure functions in a broad kinematical region.
The temperature, one of {the} free parameters of the model, is fixed to the
value $T\approx 0.06GeV$. Similar {estimates} follow also from {a}
statistical model \cite{Cleymans:1986gy, Bourrely:2001du}. Let us remark,
that {lattice QCD calculations} suggest, that the temperature corresponding
to the transition of the nuclear matter to the quark-gluon plasma is around $%
T\approx 0.175GeV$ \cite{Karsch:2001vs}. Naively one could expect, that the
average quark momenta (or temperature) in the nucleon rest frame are less
than this transition temperature. The {estimates} above do not contradict
this expectation. Further, despite {the variety of applied models, the
analysis of structure functions gives compatible} results on the measure of
intrinsic motion of quarks inside the nucleon. Roughly speaking, {the}
average momentum of the quark, if "measured" by the scattered lepton should
not exceed $\approx 0.15GeV$ in the nucleon rest frame, or $\approx 15\%$ of
the nucleon energy-momentum regardless of the reference frame. One can add,
that the leptonic information is straightforward, since after interaction
with a quark, the lepton state is not affected by other processes (final
state interaction).\medskip

\textit{B. Hadronic data (quark line)}

The Cahn effect {is a method for measuring transverse quark momenta} by
means of produced hadrons. {This} process has two stages:

1. {The} lepton-quark interaction generates {an} azimuthal asymmetry on the
level of {the} recoiled quarks, which is defined by the relations (\ref{c1}%
)--(\ref{c3}) and by the distribution of their transverse momentum.

2. {The} fragmentation of the recoiled quark {--} the asymmetry is partially
smeared in this stage. {The} inclusion of this effect requires additional
free parameters, so this method of {evaluating} the quark intrinsic motion
is less direct.

The $p_{T}$ dependence of the quark distribution function is usually
parameterized as%
\begin{equation}
f_{1}^{q}(x,p_{T})=f_{1}^{q}(x)\frac{1}{\pi \left\langle
p_{T}^{2}\right\rangle }\exp \left( -\frac{p_{T}^{2}}{\left\langle
p_{T}^{2}\right\rangle }\right) ,  \label{c20}
\end{equation}%
where%
\begin{equation}
\int \frac{1}{\pi \left\langle p_{T}^{2}\right\rangle }\exp \left( -\frac{%
p_{T}^{2}}{\left\langle p_{T}^{2}\right\rangle }\right) d^{2}p_{T}=1.
\label{c21}
\end{equation}%
One can calculate%
\begin{equation}
\left\langle p_{T}\right\rangle =\int \frac{p_{T}}{\pi \left\langle
p_{T}^{2}\right\rangle }\exp \left( -\frac{p_{T}^{2}}{\left\langle
p_{T}^{2}\right\rangle }\right) d^{2}p_{T}=\frac{\sqrt{\pi \left\langle
p_{T}^{2}\right\rangle }}{2}  \label{c22}
\end{equation}%
and from the transverse momentum {one estimates the} total momentum in the
nucleon rest frame as%
\begin{equation}
\left\langle \left\vert \mathbf{p}\right\vert \right\rangle =\sqrt{\frac{%
3\left\langle p_{T}^{2}\right\rangle }{2}}=\sqrt{\frac{6}{\pi }}\left\langle
p_{T}\right\rangle .  \label{c23}
\end{equation}

The analysis of the experimental data on the azimuthal asymmetry suggests
the following. In the paper \cite{Anselmino:2005nn} the value $\left\langle
p_{T}^{2}\right\rangle \approx 0.25GeV^{2}$ (i.e. $\left\langle
p_{T}\right\rangle \approx 0.44GeV$) is obtained (note different notation).
This result is close to the {estimate} $\left\langle p_{T}\right\rangle
\approx 0.5-0.6GeV$ following from the analyses \cite{Chay:1991nh}, \cite%
{Adams:1993hs}. Using the latest information on transverse hadron momenta
measured in semi-inclusive DIS similar numbers were obtained in an
independent approach \cite{Schweitzer:2010tt}. These figures suggest, that
the corresponding average energy-momentum of a quark in the nucleon rest
frame amounts $\approx 0.6-0.8GeV$, i.e. $\approx 64-85\%$ of the nucleon
mass. They are also substantially higher than the QCD transition temperature
mentioned above.

Obviously, two questions arise:

\textit{a)} Why do the results related to the intrinsic quark momentum
obtained by the methods \textit{A} and \textit{B}, differ by a factor
greater than four?

\textit{b)} Why does the method \textit{B} lead to a paradox, that total
energy of quarks in the nucleon rest frame can considerably exceed the
nucleon mass and related temperature is higher than the QCD transition
temperature?

We do not know the answer, but we realize, that {the} contradiction is
related to the parton model, which has its limits of validity. Nevertheless,
the questions are legitimate and require further discussion. In fact, the
inconsistency can originate in {an} arbitrary stage of the process. For
example, {the} approximation of the probability $W$ only by the one photon
exchange (\ref{c1}) can be insufficient without further QCD corrections. Or,
another function $W$ can generate {a} different degree of azimuthal
asymmetry in the general expression (\ref{d2}), which means, that fitting
the data with the false $W$ can give false $\left\langle p_{T}\right\rangle $
even though the corresponding $\chi ^{2}$ is good. Further, the quark
fragmentation into hadrons is a complex stage {containing} both pertubative
and {non-}pertubative QCD aspects. So the present {estimates} of its impact
on the smearing of primordial quark azimuthal asymmetry can be also rather
approximate. Actually, the same inconsistency is discussed also in \cite%
{D'Alesio:2009kv}.

\section{Intrinsic 3D motion in covariant parton model}

\label{sec4}This section follows from our previous study \cite{Zavada:2007ww}
and \cite{Efremov:2009ze}. In the present paper we again assume {the} quark
mass $m\rightarrow 0$. This assumption substantially simplifies {the}
calculation and seems be in a good agreement with experimental data -- in
all model relations and {sum} rules, where such {a} comparison can be done.
But in principle, {a} more complicated calculation with $m>0$ is possible 
\cite{Zavada:2002uz}. After fixing the quark mass there are no free
parameters and {the} construction of the model is based only on the two
symmetry requirements: covariance and rotational symmetry. {The} formulation
of the model in terms of the light-cone formalism is suggested in \cite%
{Efremov:2009ze} and allows to define the unpolarized leading-twist TMDs $%
f_{1}$ and $\,f_{1T}^{\perp }$ by means of the light-front correlators $\phi
(x,\mathbf{p}_{T})_{ij}$ as:%
\begin{equation}
\frac{1}{2}\;\mathrm{tr}\biggl[\gamma ^{+}\;\phi (x,\mathbf{p}_{T})\biggr]%
=f_{1}(x,\mathbf{p}_{T})-\frac{\varepsilon ^{jk}p_{T}^{j}S_{T}^{k}}{M_{N}}%
\,f_{1T}^{\perp }(x,\mathbf{p}_{T}).  \label{e1}
\end{equation}%
{The} corresponding expressions for {the} integrated {and} unintegrated
distributions $f_{1}$\ are given by Eqs. 5 and 25 in the cited paper and can
be equivalently rewritten as:%
\begin{equation}
f_{1}^{q}(x)=Mx\int G_{q}(p_{0})\delta \left( \frac{p_{0}+p_{1}}{M}-x\right) 
\frac{dp_{1}d^{2}\mathbf{p}_{T}}{p_{0}},  \label{e2}
\end{equation}%
\begin{equation}
f_{1}^{q}(x,\mathbf{p}_{T})=Mx\int G_{q}(p_{0})\delta \left( \frac{%
p_{0}+p_{1}}{M}-x\right) \frac{dp_{1}}{p_{0}}.  \label{e3}
\end{equation}%
Now we shall study these expressions in more detail. Due to rotational
symmetry in the nucleon rest frame the distribution $G_{q}$ depends on one
variable $p_{0}$; {in the} manifestly covariant representation the $p_{0}$\
is replaced by the variable $pP/M$. In this way the relation (\ref{e2})
defines {the} transformation%
\begin{equation}
G_{q}\rightarrow f_{1}^{q},  \label{q1b}
\end{equation}%
where both functions depend on one variable. In \cite{Zavada:2007ww} we
showed, that the integral (\ref{e2}) can be inverted%
\begin{equation}
G_{q}(p)=-\frac{1}{\pi M^{3}}\left( \frac{f_{1}^{q}(x)}{x}\right) ^{\prime },
\label{q1a}
\end{equation}%
where%
\begin{equation*}
x=\frac{2p}{M},\quad p\equiv p_{0}=\sqrt{p_{1}^{2}+p_{T}^{2}}.
\end{equation*}%
In this way the distributions $G_{q}$\ can be obtained from {the}
distributions $f_{1}^{q}$, which are extracted from the structure functions
by global analysis.\ Apparently, there is {a} one-to-one mapping%
\begin{equation}
G_{q}(p)\leftrightarrows f_{1}^{q}(x)  \label{q1c}
\end{equation}%
so both distributions represent equivalent descriptions.

Now, we will calculate the TMD integral (\ref{e3}). First we calculate roots
of the expression in the \ $\delta -$ function for the variable $p_{1}$:%
\begin{equation}
\frac{p_{0}+p_{1}}{M}-x=0,  \label{q2}
\end{equation}%
there is just one root%
\begin{equation}
\tilde{p}_{1}=\frac{Mx}{2}\left( 1-\left( \frac{p_{T}}{Mx}\right)
^{2}\right) .  \label{q3}
\end{equation}%
At the same time the corresponding variable \ $p_{0}$\ \ reads:%
\begin{equation}
\tilde{p}_{0}=\frac{Mx}{2}\left( 1+\left( \frac{p_{T}}{Mx}\right)
^{2}\right) .  \label{q4}
\end{equation}%
The $\delta -$function term can be modified as 
\begin{eqnarray}
\delta \left( \frac{p_{0}+p_{1}}{M}-x\right) dp_{1} &=&\frac{\delta \left(
p_{1}-\tilde{p}_{1}\right) dp_{1}}{\left\vert \frac{d}{dp_{1}}\left( \frac{%
p_{0}+p_{1}}{M}-x\right) _{p_{1}=\tilde{p}_{1}}\right\vert }  \notag \\
&=&\frac{\delta \left( p_{1}-\tilde{p}_{1}\right) dp_{1}}{x/p_{0}},
\label{q5}
\end{eqnarray}%
then after inserting to Eq. (\ref{e3}) one gets:%
\begin{equation}
f_{1}^{q}(x,\mathbf{p}_{T})=M\int G_{q}(p_{0})\delta \left( p_{1}-\tilde{p}%
_{1}\right) dp_{1}=MG_{q}(\tilde{p}_{0}).  \label{q6}
\end{equation}%
One can observe, that $f_{1}^{q}(x,\mathbf{p}_{T})$ depends on $x,\mathbf{p}%
_{T}$ via one variable $\tilde{p}_{0}$ defined by Eq. (\ref{q4}). It is due
to fact, that this variable in $G_{q}(\tilde{p}_{0})$ reflects rotational
symmetry in the rest frame. Obviously $x,\mathbf{p}_{T}$ are not independent
variables at fixed $p_{0}$ or $p_{1}$. Also in Eq. (\ref{q6}) both functions
represent equivalent description. Further, if we define%
\begin{equation}
\xi =x\left( 1+\left( \frac{p_{T}}{Mx}\right) ^{2}\right) ,  \label{q8}
\end{equation}%
then 
\begin{equation}
f_{1}^{q}(x,\mathbf{p}_{T})=MG_{q}\left( \frac{M}{2}\xi \right) .  \label{q9}
\end{equation}%
Since Eq. (\ref{q1a}) implies%
\begin{equation}
G_{q}\left( \frac{M}{2}\xi \right) =-\frac{1}{\pi M^{3}}\left( \frac{%
f_{1}^{q}(\xi )}{\xi }\right) ^{\prime },  \label{q10}
\end{equation}%
after inserting to Eq. (\ref{q9}) we get the result%
\begin{equation}
f_{1}^{q}(x,\mathbf{p}_{T})=-\frac{1}{\pi M^{2}}\left( \frac{f_{1}^{q}(\xi )%
}{\xi }\right) ^{\prime };\ \xi =x\left( 1+\left( \frac{p_{T}}{Mx}\right)
^{2}\right) .  \label{q11}
\end{equation}%
This equation represents new {relation}, which connects integrated and
unintegrated unpolarized distribution functions. Before further discussion
one can verify {the} compatibility with Eqs. (\ref{e2}) and (\ref{e3}):%
\begin{equation}
f_{1}^{q}(x)=\int f_{1}^{q}(x,\mathbf{p}_{T})d^{2}\mathbf{p}_{T}.
\label{q12}
\end{equation}%
Eq. (\ref{q11}) implies%
\begin{equation}
\int f_{1}^{q}(x,\mathbf{p}_{T})d^{2}\mathbf{p}_{T}=-\frac{2}{M^{2}}%
\int_{0}^{p_{T\max }(x)}\left( \frac{f_{1}^{q}(\xi )}{\xi }\right) ^{\prime
}p_{T}dp_{T},  \label{q13}
\end{equation}%
where we replaced $d^{2}\mathbf{p}_{T}=2\pi p_{T}dp_{T}.$ From Eq. (\ref{q8}%
) we have%
\begin{equation}
d\xi =\frac{2p_{T}dp_{T}}{M^{2}x}.  \label{q14}
\end{equation}%
and Eqs. (\ref{c18}) and (\ref{q8}) imply 
\begin{equation}
x\leq \xi \leq 1.  \label{q16}
\end{equation}%
Now the Eq. (\ref{q13}) can be modified as%
\begin{equation}
\int f_{1}^{q}(x,\mathbf{p}_{T})d^{2}\mathbf{p}_{T}=-x\int_{x}^{1}\left( 
\frac{f_{1}^{q}(\xi )}{\xi }\right) ^{\prime }d\xi ,  \label{q17}
\end{equation}%
from which Eq. (\ref{q12}) follows easily. 
\begin{figure}[tbp]
\includegraphics[width=9cm]{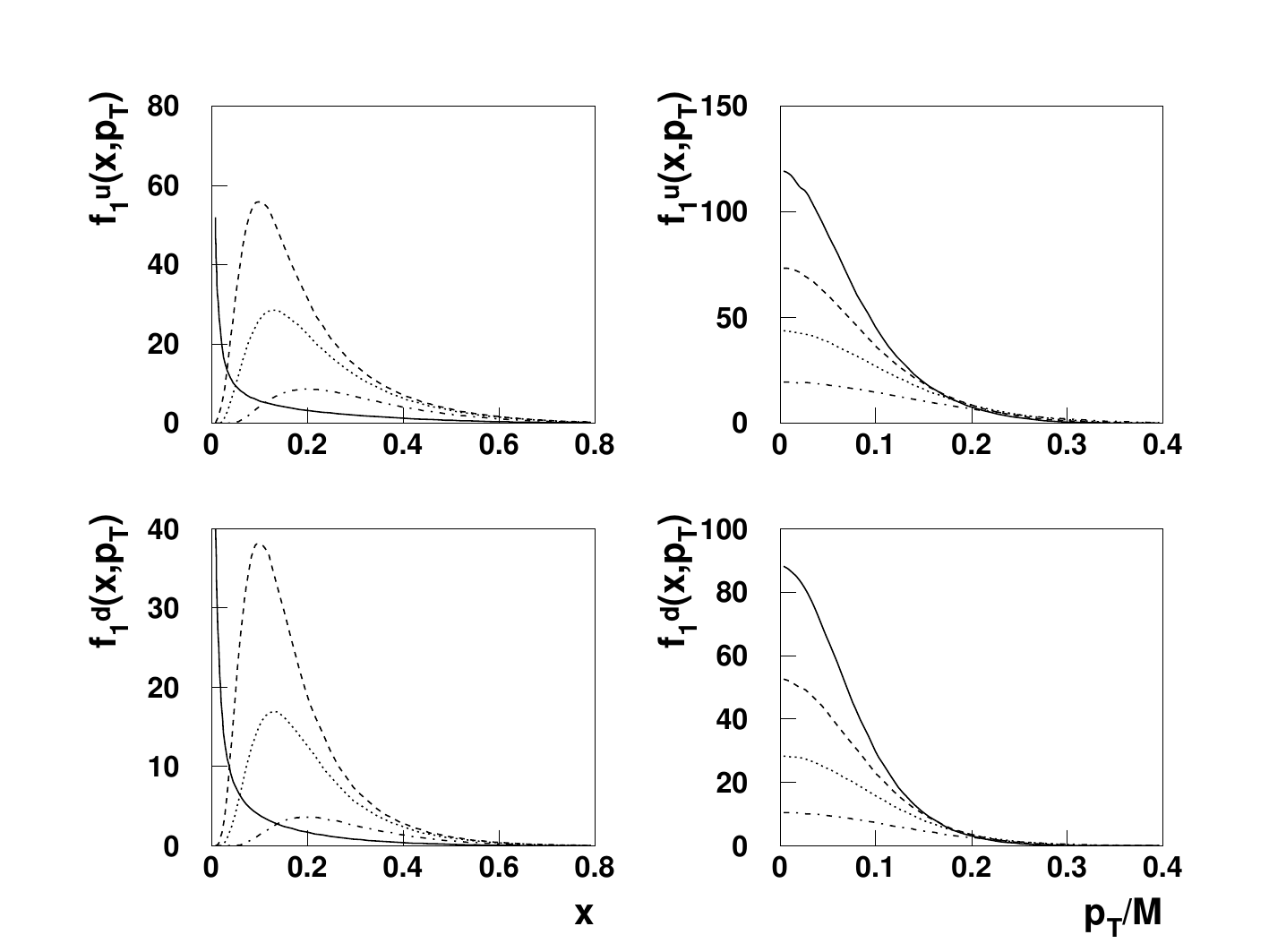}
\caption{Transverse momentum dependent unpolarized distribution functions
for $u$ (\textit{upper panel}) and $d-$quarks (\textit{lower panel}). 
\textbf{Left panel}: dependence on $x$ for $p_{T}/M=0.10,0.13,0.20$ is
indicated by dash, dotted and dash-dot curves; solid curve correspods to the
integrated distribution $f_{q}(x)$. \textbf{Right panel}: dependence on $%
p_{T}/M$ $\ $for $x=0.15,0.18,0.22,0.30$ is indicated by solid, dash, dotted
and dash-dot curves.}
\label{fg2}
\end{figure}

We can make two remarks about the obtained results:

\textit{i)} Due to covariance and rotational symmetry (which follows from {%
the invariant variable} $pP/M$ in the rest frame), all the following
distributions used in our approach involve equivalent information 
\begin{equation}
f_{1}^{q}(x,\mathbf{p}_{T})\Leftrightarrow G_{q}(\mathbf{p})\Leftrightarrow
G_{q}(p_{0})\Leftrightarrow G_{q}(\frac{pP}{M})\Leftrightarrow f_{1}^{q}(x)
\label{qq18}
\end{equation}%
and also the two sets of variables are equivalent:%
\begin{equation}
\mathbf{p}\Leftrightarrow x,\mathbf{p}_{T},\qquad d^{3}\mathbf{p=}\frac{p_{0}%
}{x}dxd^{2}\mathbf{p}_{T}.  \label{q19}
\end{equation}

\textit{ii)} All the functions (\ref{qq18}) are assumed to depend also on $%
Q^{2}$, although the evolution is not involved in the present version of \ {%
the} model. Nevertheless, due to this equivalence, in {the \textsl{present}}
approach the evolution of $f_{1}^{q}(x,\mathbf{p}_{T},Q^{2})$ can be
obtained from $f_{1}^{q}(x,Q^{2})$, which is {evolved in} standard way and
similarly for the distribution $G_{q}(\mathbf{p},Q^{2})$.

Now, we can apply the obtained relations for corresponding numerical
calculation. {The} transverse momentum {dependent} distribution functions $%
f_{1}^{q}(x,\mathbf{p}_{T})$ are calculated from Eq. (\ref{q11}), for input
distributions $f_{1}^{q}(x)$ we used the standard parameterization \cite%
{Martin:2004dh} (LO at the scale $4GeV^{2}$).\ In Fig. \ref{fg2} we have
results for $u$ and $d-$quarks. {The left panel} demonstrates, that $x$ and $%
p_{T}$ are not independent variables. In accordance with the relation (\ref%
{c18}), in the sample of partons with fixed $p_{T}$ the region of low $x$ is
effectively suppressed. For {larger} $p_{T}$ the effect is getting more
pronounced. The right {panel} of the figure demonstrates, that {the} typical
value of $p_{T}$ in this approach corresponds to the estimates based on the
leptonic data in Sec. \ref{sec3}.

Further, let us compare our model giving the {relation} (\ref{q11}) with the
approach described in the recent paper \cite{D'Alesio:2009kv}. The
corresponding relation (57) in the cited paper reads:%
\begin{equation}
q(x,\mathbf{k}_{T}^{2})=-\frac{1}{\pi M^{2}}\frac{d}{dx}\left[ \frac{q(x)}{x}%
\right] _{x=\eta }\theta \left[ x\left( 1-x\right) M^{2}-\mathbf{k}_{T}^{2}%
\right] ,  \label{eliot1}
\end{equation}%
where%
\begin{equation*}
\eta =x+\frac{\mathbf{k}_{T}^{2}}{xM^{2}}. 
\end{equation*}%
We agree with the authors of cited paper, that both relations are equivalent
({the} authors refer to a first version of our paper). The $\theta $%
--function term corresponds to the constraint (\ref{c18}) valid in our
approach and the correspondence of other symbols is obvious. The relation (%
\ref{eliot1}) follows from a previous relation (55) in \cite{D'Alesio:2009kv}%
\begin{equation}
q(x,\mathbf{k}_{T}^{2})=\frac{1}{\pi M^{2}}\varphi _{3}\left( x+\frac{%
\mathbf{k}_{T}^{2}}{xM^{2}}\right) \theta \left[ x\left( 1-x\right) M^{2}-%
\mathbf{k}_{T}^{2}\right] .  \label{eliot2}
\end{equation}%
If we integrate this equation, then the l.h.s. represents the definition
(54) in \cite{D'Alesio:2009kv}%
\begin{equation}
\int q(x,\mathbf{k}_{T}^{2})d^{2}\mathbf{k}_{T}=q(x)  \label{eliot3}
\end{equation}%
and after substitutions $d^{2}\mathbf{k}_{T}\rightarrow \pi d\mathbf{k}%
_{T}^{2}$, $\mathbf{k}_{T}^{2}\rightarrow \eta =x+\mathbf{k}_{T}^{2}/xM^{2}$
the r.h.s. gives%
\begin{equation}
\frac{1}{\pi M^{2}}\int_{0}^{x\left( 1-x\right) M^{2}}\varphi _{3}\left( x+%
\frac{\mathbf{k}_{T}^{2}}{xM^{2}}\right) d^{2}\mathbf{k}_{T}=x\int_{x}^{1}%
\varphi _{3}\left( \eta \right) d\eta .  \label{eliot4}
\end{equation}%
The last two equations imply the relation%
\begin{equation}
q(x)=x\int_{x}^{1}\varphi _{3}\left( \eta \right) d\eta ,  \label{eliot5}
\end{equation}%
which after differentiation and inserting into Eq. (\ref{eliot2}) gives the
final relation (\ref{eliot1}). The approach developed in \cite%
{D'Alesio:2009kv} is motivated by the classic papers \cite{Ellis:1982wd}, 
\cite{Ellis:1982cd} {from} which also the starting equation (\ref{eliot2})
is adopted. The corresponding model is represented by the handbag diagram,
in which the incoming line is put on-mass-shell $k^{2}=0$ but has non-zero
transverse momentum \cite{Ellis:1982wd}, fig. 1a. Let us also remark, that {%
the} form of the expression (\ref{eliot2}) is dictated by Lorentz
invariance. Further, comparing this expression with Eq. (\ref{q9}) allows to
identify 
\begin{equation}
\frac{1}{\pi M^{3}}\varphi _{3}\left( \xi \right) =G_{q}\left( \frac{M}{2}%
\xi \right) ,  \label{eliot6}
\end{equation}%
where%
\begin{equation}
\xi =x+\frac{\mathbf{p}_{T}^{2}}{xM^{2}}=\frac{2P\cdot p}{M^{2}},
\label{eliot7}
\end{equation}%
see e.g. Eq. (28) in \cite{D'Alesio:2009kv}. The last equality means, that
in the nucleon rest frame $\xi =2p_{0}/M$, which implies rotational symmetry
of the both functions $\varphi _{3}$ and $G_{q}$ in this frame.

So we can conclude, that both approaches have {a} common basis represented
by {the} requirements:

\textit{i)} Lorentz invariance, which in fact implies also rotational
symmetry of the quark momentum distribution in the nucleon rest frame.

\textit{ii)} quarks are on-mass-shell: $p^{2}=0.$

The equivalent results, like Eqs. (\ref{q11}) and (\ref{eliot1}) are just {a}
consequence of these conditions. {The Wandzura}-Wilczek relation obtained
equally in \cite{D'Alesio:2009kv} and \cite{Zavada:2001bq} is a further
example. At the same time it is apparent, that despite {a} common input, {the%
} procedures applied in both approaches are substantially different. Other
distribution functions like transversity or pretzelosity require additional
assumptions {to be included in} the approach, so the corresponding results
from both approaches may differ depending on the chosen method of
generalization.

\section{Summary and conclusion}

\label{sec5}We studied some questions related to the distribution of quark
transverse momenta in the framework of {the} covariant approach. From this
point of view, this distribution is a projection of {a} more general 3D
motion of quarks inside the nucleon {with respect} to the plane transverse
to the momentum of {the} probing particle. Due to general arguments, {the}
3D motion of quarks in the nucleon rest frame has rotational symmetry. We
suggested, that in our approach this rotational symmetry follows from
covariance (Lorentz invariance). It follows, that {in both} pictures 2D and
3D momenta distributions involve equivalent information. The main results
obtained in this paper can be summarized as follows.

\textit{i)} We analyzed {the} conditions generating the Cahn effect, which
represents {an} important tool for measuring of the quark transverse motion.
We suggested, that the effect has a more general origin than it is currently
considered. We obtained {a} general expression for azimuthal asymmetry,
which depends on intrinsic transverse momentum of the quarks and on the
probability $W(\hat{s},\hat{u})$ of \ the lepton-quark scattering. At the
same time we presented arguments, why the analysis of data on azimuthal
asymmetry due to Cahn effect requires caution.

\textit{ii)} We have done a comparison, which suggests that the data on
transverse motion based on Cahn effect disagree with the data based on
analysis of the structure functions ($F_{2}$) in the framework of various
models. Both methods differ in estimation of $\left\langle
p_{T}\right\rangle $\ by factor $\approx 4$.

\textit{iii)} We studied {the} unpolarized parton distribution functions $%
f_{1}^{q}(x,\mathbf{p}_{T})$ in the framework of the 3D covariant parton
model. We obtained a new {relation}, which relates this TMD to its
integrated counterpart $f_{1}^{q}(x)$. Using this relation with the input on 
{the} integrated distribution obtained from global analysis, we calculated $%
f_{1}^{q}(x,\mathbf{p}_{T})$\ also numerically.

\textit{iv)} We have done a detailed comparison with the recent approach\ by
U.D'Alesio, E.Leader and F.Murgia \cite{D'Alesio:2009kv}, in which an
equivalent {relation and other} results coincident with our approach have
been obtained. We have proved, that both approaches have {a} common general
basis consisting in Lorentz invariance (covariance) and in the on-mass-shell
condition $p^{2}=0.$ That is why, despite substantially different procedures
and formalism applied in both approaches, some results are identical.

\textit{v)} We confirmed, that the requirement of relativistic covariance
combined with the nucleon rotational symmetry {represents a} powerful tool
for revealing new {relations} connecting various parton distribution
functions, including the relations between the integrated (PDF) and their
unintegrated (TMD) counterparts.

\begin{acknowledgments}
This work was supported by the project AV0Z10100502 of the Academy of
Sciences of the Czech Republic. I am grateful to Anatoli Efremov, Peter
Schweitzer and Oleg Teryaev for many useful discussions and valuable
comments. I would like to thank also to Jacques Soffer and Claude Bourrely
for helpful comments on earlier version of the manuscript.
\end{acknowledgments}

\end{document}